\documentclass[11pt]{article}
\usepackage{moriond2000,epsfig}

\def\Journal#1#2#3#4{{#1} {\bf #2}, #3 (#4)}

\def\be{\begin{equation}}
\def\ee{\end{equation}}
\def\bea{\begin{eqnarray}}
\def\eea{\end{eqnarray}}

\begin{document}
\vspace*{4cm}
\title{THE CURRENT STATUS OF CLASS}

\author{ P. HELBIG (for the CLASS collaboration) }

\address{Rijksuniversiteit Groningen, Kapteyn Instituut, Postbus 800,\\
NL-9700 AV Groningen, The Netherlands}

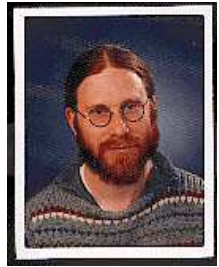
\maketitle\abstracts{
I give a brief overview of the current status of some aspects of the 
Cosmic Lens All-Sky Survey, CLASS.
}

\section{Who and What?}

CLASS, the Cosmic Lens All-Sky Survey, is a collaboration between groups
at the Jodrell Bank Observatory (UK), ASTRON (NL), the Kapteyn
Astronomical Institute at the University of Groningen (NL), Caltech
(USA), the University of Pennsylvania (USA) and NRAO (USA).  It is a 
survey for flat-spectrum radio sources ($\alpha \ge -0.5$ for $S_{f} \sim
f^{\alpha}$ between L-band (Condon et al.\cite{NVSS} (NVSS)) and C-band 
(Gregory et al.\cite{GB6} (GB6))) with the selection criteria 
$0 \le \delta \le 75^{\circ}$ (due to the area covered by GB6),
$|b| > 10^{\circ}$ and $S_{\mathrm{C-band}} > 30$\,mJy: 11685 objects.  The 
objects thus selected were then observed with the VLA in A-configuration 
at X-band.  CLASS has many goals, many of which are lens-related 
(measuring $\lambda_{0}$ and $\Omega_{0}$ with lensing statistics and 
$H_{0}$ with time delays, studying the properties of lensing galaxies).
(The definition of CLASS has evolved, mainly due to improvements in the 
input catalogues used.  The above definition is the current, final one.
However, objects were observed which are not part of the currently 
definied `statistically complete CLASS' and, if they happened to be lens 
systems, were of course followed up.)

\section{Why and How?}

Radio surveys can be particularly good gravitational-lens surveys, for a 
number of reasons:
\begin{itemize}
\item Using interferometry, the beam size is much smaller than the image 
separation.
\item Flat-spectrum objects are compact (i.e. (almost) point sources), 
allowing typical lensing morphologies to be recognised easily.
\item Somewhat related to the previous point, the lensing probability 
depends on the cross section determined by the lens population; for 
extended sources, the source geometry partially determines what is 
recognised as a lens system.
\item Most sources are quasars at high redshift, which leads to a high 
lensing rate.
\item Since flat-spectrum objects are compact, they can be variable on 
relatively short timescales, which aids in determining time delays.
\item There is no bias from lens galaxies due to extinction by the lens or 
comparable brightness of source and lens, as can be the case with 
optical surveys.
\item High-resolution followup is possible with interferometers such as 
MERLIN, the VLBA, VLBI\dots.
\end{itemize}
But there is one disadvantage:
\begin{itemize}
\item Additional work is required to get redshifts.
\end{itemize}

From our own CLASS VLA observations (i.e.\ A-configuration at X-band),
objects with multiple compact ($<170$\,mas) are marked as lens
candidates if the separation is between 300\,mas and 6\,arcsec, the flux
ratio $<$10:1 and the total flux at X-band in the compact components is
$>$20\,mJy.  These constraints are chosen so that sufficient followup is
possible for all such candidates.  Candidates are then observed at
progressively higher resolution, first at C-band with MERLIN (50\,mas
resolution) and then, if necessary, the VLBA (3\,mas resolution) and 
finally if needed with VLBI, at additional frequencies if needed.  
About 80\% of the candidates are rejected after the MERLIN `filter' due 
to different surface brightnesses in the components (gravitational 
lensing, of course, conserves surface brightness).  Other reasons for 
rejection include obvious non-lens structure (such as a core-jet 
morphology) and different spectral indices or polarisation.  It is 
interesting to note that, with one possible exception, every candidate 
which has passed these tests and thus been deemed to be a lens system on the 
basis of radio data alone has yielded a lens galaxy when observed 
optically (usually with HST).  Since the source and lens redshifts are 
needed for various purposes, these are also checked to be consistent 
with lensing ($z_{\mathrm{l}} < z_{\mathrm{s}}$, the same 
$z_{\mathrm{s}}$ for all components of the source), though in practice 
no candidates which have passed the above-mentioned tests have been 
ruled out on this basis.

Table~1 shows the current CLASS and JVAS lens systems.
\begin{table*}[h!]
\caption[]{The JVAS and CLASS gravitational lenses}
\begin{tabular*}{\textwidth}{@{\extracolsep{\fill}}|lllllll|}
\hline
Survey &
Name & 
\# images & 
$\Delta\theta''$ & 
$z_{\mathrm{l}}$ & 
$z_{\mathrm{s}}$ & 
lens galaxy \\
\hline
\multicolumn{7}{|c|}{\textbf{confirmed lenses}} \\
CLASS         &                
B0128+437     &                
4             &                
0.542         &                
\textbf{?}    &                
\textbf{?}    &                
\textbf{?}    \\               
JVAS          &                
B0218+357     &                
2 + ring      &                
0.334         &                
0.6847        &                
0.96          &                
spiral        \\               
JVAS          &                
MG0414+054    &                
4             &                
2.09          &                
0.9584        &                
2.639         &                
elliptical    \\               
CLASS         &                
B0712+472     &                
4             &                
1.27          &                
0.406         &                
1.34          &                
spiral        \\               
CLASS         &                
B0739+366     &                
2             &                
0.540         &                
\textbf{?}    &                
\textbf{?}    &                
\textbf{?}    \\               
JVAS          &                
B1030+074     &                
2             &                
1.56          &                
0.599         &                
1.535         &                
spiral        \\               
CLASS         &                
B1127+385     &                
2             &                
0.701         &                
\textbf{?}    &                
\textbf{?}    &                
\textbf{?}    \\               
CLASS         &                
B1152+119     &                
2             &                
1.56          &                
0.439         &                
1.019         &                
\textbf{?}    \\               
CLASS         &                
B1359+154     &                
4             &                
1.65          &                
\textbf{?}    &                
3.212         &                
\textbf{?}    \\               
JVAS          &                
B1422+231     &                
4             &                
1.28          &                
0.337         &                
3.62          &                
\textbf{?}    \\               
CLASS         &                
B1555+375     &                
4             &                
0.43          &                
\textbf{?}    &                
\textbf{?}    &                
\textbf{?}    \\               
CLASS         &                
B1600+434     &                
2             &                
1.39          &                
0.414         &                
1.589         &                
spiral        \\               
CLASS         &                
B1608+656     &                
4             &                
2.08          &                
0.63          &                
1.39          &                
spiral        \\               
CLASS         &                
B1933+507     &                
$4+4+2$       &                
1.17          &                
0.755         &                
2.62          &                
\textbf{?}    \\               
JVAS          &                
B1938+666     &                
$4+2$         &                
0.93          &                
0.878         &                
\textbf{?}    &                
\textbf{?}    \\               
CLASS         &                
B2045+265     &                
4             &                
1.86          &                
0.867         &                
1.28          &                
\textbf{?}    \\               
CLASS         &                
B2319+051     &                
2             &                
1.365         &                
0.624         &                
\textbf{?}    &                
\textbf{?}    \\               
\hline
\multicolumn{7}{|c|}{\textbf{puzzling probable lenses}} \\
JVAS          &                
B2114+022     &                
2 or 4        &                
2.57          &                
0.32 \& 0.59  &                
\textbf{?}    &                
\textbf{?}    \\               
\hline
\end{tabular*}
\end{table*}
It should be noted that missing redshifts are mostly from relatively new 
systems (though in the case of B1938+666, a redshift might could have been 
measured if UKIRT were able look this far north).  It thus appears to be a 
realistic goal to have the survey complete with respect to both source 
and lens redshifts within the next several months.  Note also the 
relatively narrow range of image separations, which is definitively 
\emph{not} due to any sort of bias or incompleteness but reflects a real 
fact about the universe (more precisely, the mass spectrum of lens 
galaxies).

\section{$\lambda_{0}$ and $\Omega_{0}$ from Lensing Statistics}

We have done an analysis of JVAS (essentially the brightest 2308 sources 
in CLASS).  Lensing statistics, at least in the interesting part of 
parameter space, essentially measures $\lambda_{0}-\Omega_{0}$.  As 
discussed in Helbig\cite{JVAS}, we obtain 
\begin{equation}
-2.69 < \lambda_{0} - \Omega_{0} < +0.68
\end{equation}
at 95\% confidence; for a flat universe, this corresponds to
\begin{equation}
-0.85 < \lambda_{0} - \Omega_{0} < +0.84
\end{equation}
It should be noted that the probability distribution in the 
$\lambda_{0}$-$\Omega_{0}$ plane is not Gaussian; in particular, the 
above numbers were obtained from `real contours' (cf.\ Helbig\cite{SNIa} 
and references therein) and not by plotting contours at some fraction of 
the peak likelihood, and of course they depend on the region of 
parameter space examined as long as there is a non-negligible likelihood 
outside of it.  These caveats should be kept in mind when comparing 
these constraints to others in the literature.

Of course, not only an analysis of CLASS but a better analysis of CLASS 
is in the works.  However, first the survey and lens followup have to be 
finished and the $S$-$z$ plane of the parent population (i.e.\ 
non-lenses in the survey and objects `amplified in' to the survey by 
lensing) must be determined.  This information is needed in two quite 
distinct contexts, even though they are really two sides of the same coin:
\begin{itemize}
\item The number--flux-density relation at the redshifts of the sources 
in the lens systems is needed for the calculation of the amplification 
bias.
\item The flux-density--dependent redshift distribution is needed as a 
proxy for the redshifts of the non-lenses in the survey.
\end{itemize}
Even without this information, however, there is the interesting 
possibility of using CLASS for the lens-redshift test; see 
Helbig\cite{zl} for discussion.

\section{$H_{0}$ from Time Delays} 

Since all observables in a gravitational lens system are dimensionless
except for the time delay between the images, this can be used to scale
the model of the system and thus, if the redshifts are known, measure
$H_{0}$ (with a higher-order dependency on $\lambda_{0}$ and
$\Omega_{0}$).  A radio survey offers the advantages that microlensing
is less of a worry and that the sources are `pre-selected' to be
variable (due to their flat-spectrum nature, as mentioned above).  The
angular separation probed by CLASS corresponds to galaxy-mass lenses and
thus time delays of weeks, which is convenient.  To date, there are 6
measured gravitational-lens time delays, 3 from CLASS, of which 5 agree
quite well (see, e.g., Koopmans \& Fassnacht\cite{KF} for discussion).
In principle, as first pointed out by Refsdal\cite{SR}, one can use time
delays from several systems to measure $\lambda_{0}$ and $\Omega_{0}$.
Although it is too early to make a definitive statement, it is
interesting to note that the derived values for $H_{0}$ agree better if
currently favoured values for $\lambda_{0}$ and $\Omega_{0}$ are
assumed.  To quote a number for posterity, $H_{0}=68$.  More CLASS 
systems are being monitored, so statistics should improve in the future.

\section{Dark Lenses?}

As mentioned above, there is only one possible lens system which has
passed all the radio tests but in which no lens galaxy has been
detected, and could thus be a `dark lens' (cf.\ Jackson et 
al.\cite{dark} and 
references therein).  However, statistical arguments favour the 
alternative explanation of this system, B0827+525, being the first 
binary radio-loud QSO.  On the other hand, the case is not clear-cut.
If the latter explanation is true, it will have the smallest separation 
of all known QSO pairs.  On the other hand, if it is a lens system, it 
will have the largest separation of all CLASS lenses.  See Koopmans et 
al.\cite{0827} for further discussion.

\section{Wide-Separation Lenses}

Although CLASS originally examined the range between 300\,mas and
6\,arcsec, this has been extended up to an arc-minute in two new
surveys.  First, CLASS has been extended to search up to 15\,arcsec in a
manner identical to the original survey, made possible by the increase
in computing power since CLASS began.  Second, a new survey, the
Arc-Minute Radio Cluster-Lens Search (ARCS), has looked for lensing of 
extended sources on the 15 to 60 arcsec scale (due to the longer time 
delays, many of the arguments used to rule out candidates at smaller 
separation will not work at larger separation and the number of chance 
coincidences on the sky of course increases with the 
separation, so a different strategy is called for).  To date, \emph{no}
lens systems have been found in this range of angular separation, though 
one candidate remains.  More details can be found in Phillips et 
al.\cite{ARCS}.

\section*{Acknowledgments}

It is a pleasure to thank the rest of the CLASS collaboration for input.


\section*{References}
\begin{thebibliography}{99}

\bibitem{NVSS}J.J.~Condon et al., \Journal{AJ}{115}{1693}{1998}

\bibitem{GB6}P.C.~Gregory et al.,\Journal{ApJS}{103}{427}{1996}

\bibitem{JVAS}P.~Helbig et al., \Journal{A\&A}{136}{297}{1999}

\bibitem{SNIa}P.~Helbig \Journal{A\&A}{350}{1}{1999}

\bibitem{zl}P.~Helbig, these proceedings

\bibitem{KF}L.V.E.~Koopmans \& C.D.~Fassnacht, \Journal{ApJ}{527}{513}{1999}

\bibitem{SR}S.~Refsdal, \Journal{MNRAS}{132}{101}{1966}

\bibitem{dark}N.~Jackson et al., \Journal{A\&A}{334}{L33}{1998}

\bibitem{0827}L.~Koopmans et al., A\&A, in press

\bibitem{ARCS}P.~Phillips et al., MNRAS, submitted

\end{thebibliography}
\end{document}